# AN EFFICIENT DOMAIN-INDEPENDENT APPROACH FOR SUPERVISED KEYPHRASE EXTRACTION AND RANKING

Sriraghavendra Ramaswamy

Amazon.com
sriragha@amazon.com

## ABSTRACT

*We present a supervised learning approach for automatic extraction of keyphrases from single documents. Our solution uses simple to compute statistical and positional features of candidate phrases and does not rely on any external knowledge base or on pre-trained language models or word embeddings. The ranking component of our proposed solution is a fairly lightweight ensemble model. Evaluation on benchmark datasets shows that our approach achieves significantly higher accuracy than several state-of-the-art baseline models, including all deep learning-based unsupervised models compared with, and is competitive with some supervised deep learning-based models too. Despite the supervised nature of our solution, the fact that does not rely on any corpus of "golden" keywords or any external knowledge corpus means that our solution bears the advantages of unsupervised solutions to a fair extent.*

## 1. INTRODUCTION

Keyphrase prediction is the process of finding a small set of phrases that best represent the main text content of a document. It plays an important role in search and information retrieval by providing a convenient way to index, classify, organize and summarize documents [1, 2]. Keyphrases can either be "present keyphrases" that occur as such in the document, or "absent keyphrases" that do not match any word sequence in the text but represent a topic term or a concept discussed in the document [3]. In this paper, we tackle the problem of automatic selection of a good set of "present" keyphrases, a task often referred to as AKE or PKE (automatic/present keyphrase extraction), or simply as keyphrase extraction.

The recent years have witnessed a shift towards using deep learning (DL) to solve keyphrase extraction, often by posing it as a variant of keyword generation (predicting "absent" keyphrases) task [3]. Various studies [2, 3] report that DL-based extraction methods outperform traditional unsupervised methods on accuracy metrics. Yet, while the indispensability of DL methods for high quality keyphrase generation is evident, their optimality for purely keyphrase extraction tasks is debatable. Not all applications needing keyword extraction will benefit from say a 0.05 increase in F1-score coming at the high cost of LLMs. Some well-known non-DL supervised solutions for AKE do achieve accuracy levels close to those of DL-based techniques, but these solutions are implicitly domain-specific in their applicability.

In this paper, we devise a lightweight supervised machine learning approach for automatic keyphrase extraction based predominantly on simple to compute statistical and positional features. Unlike prior known non-DL approaches that reach similar accuracy levels, our method does not depend on any knowledge base (barring phrase frequency corpus created from training set), be it semantic relationship knowledge graphs, pre-computed topic models or word embeddings or a corpus of "golden" keywords (a.k.a. seed keywords or core keywords). This makes our approach fairly domain agnostic and capable of generalizing better for documents pertaining to a wider spectrum of subjects/domains. We evaluate our model on widely used benchmark datasets. Our model surpasses the accuracy of several benchmark models, including some deep learning models on the two benchmark datasets we used for evaluation. In one these datasets, our model's accuracy is quite close to the levels achieved by an LLM-based solution.

## 2. RELATED WORK

Automatic keyphrase extraction methods are broadly classified into two types — unsupervised and supervised. In unsupervised methods, AKE is treated as a ranking problem. Since there is no access to annotated data in these methods, candidate ranking is done using scores computed heuristically. Unsupervised methods are further classified into statistical methods and graph-based methods based on the types of features used to compute the scores. Statistical approaches, for e.g. TF-IDF [4] and YAKE [5], use statistical features such as word frequencies and cooccurrence counts. In contrast, graph-based approaches (e.g., TextRank [6]) construct a graph representation of text, such as with words serving as nodes and their co-occurrences as edges. Thereafter, a node ranking algorithm (e.g., PageRank) is used to sort words/phrases, and return the top k candidate keyphrases.

In supervised methods, a classifier is trained on documents annotated with keyphrases and this classifier is used on new documents to determine whether a candidate phrase is a keyphrase or not. One of the first keyphrase extraction methods, KEA [7], trains a Naive Bayes classifier with TF-IDF score and the first occurrence position used as features of a candidate phrase.

Both supervised and unsupervised methods may use external knowledge for candidate selection and ranking. For e.g, Maui [8], a supervised method, computes a keyphraseness feature based on how often a candidate phrase appears as a keyphrase in the training corpus (Note: this is not to be confused with document frequency (DF) of a phrase in the training corpus; DF does not need knowledge of whether or not the phrase is a keyphrase in the training corpus). KeyCluster [9], an unsupervised method, uses Wikipedia data to compute term relatedness.

In the recent years, deep learning has become the preferred mechanism for keyphrase prediction solutions. These solutions can be meant either solely for keyphrase extraction (e.g., EmbedRank [10], UKERank [11]) or for absent/abstractive keyphrase generation too (e.g., CopyRNN [12]).

## 3. OUR APPROACH

In this and the next section, we describe our novel keyphrase extraction method. The method consists of three main steps:

(1) Extract candidate phrases from the document text based on part-of-speech sequences

(2) Compute features for every candidate phrase

(3) Rank the candidate phrases using either a partial ranking model or a classification model that uses the features from step 2. The model gives a score for every candidate phrase and we select the top k keyphrases, where k is the desired number of keyphrases to be selected from every single input document.

### 3.1. Problem Formulation

We frame keyphrase selection from candidate phrases as a problem of partial ranking. We want to find a scoring function $\mathcal{F}$ such that given an input document $\mathcal{D}$ with $|\mathcal{P}|$ distinct candidate phrases $\mathcal{P} = \{p_1, p_2, …, p_{|\mathcal{P}|}\}$, the function $\mathcal{F} : \mathcal{P} \mapsto \mathbb{R}$ is such that $\mathcal{F}(p_i) \geq \mathcal{F}(p_j) \Leftrightarrow \mathbf{1}_{\mathcal{K}_\mathcal{D}}(p_I) \geq \mathbf{1}_{\mathcal{K}_\mathcal{D}}(p_j)$ where $\mathbf{1}_{\mathcal{K}_\mathcal{D}}(.)$ is the indicator function with $\mathcal{K}_\mathcal{D}$ being the "ideal" set of keyphrases for document $\mathcal{D}$. In other words, if a phrase $p_a$ gets a higher score than phrase $p_b$ then $p_a$ must be equally or more "ideal" compared to $p_b$, i.e. it cannot be the case that $p_a$ is not an "ideal" keyphrase than $p_b$ whereas is. Or more simply put, ideal keyphrases should not get a lower score than any non-keyphrase. $\mathcal{K}_\mathcal{D}$ is of course not known a priori (except when the document belongs to the training set) — inferring $\mathcal{K}_\mathcal{D}$ is the goal of our model. During the training phase we use labelled datasets wherein keyphrases assigned manually (by authors, reviewers, or expert readers) to each document are known; we treat those as the ideal keyphrases for the corresponding documents and this forms a basis of our supervised learning solution.

We solve the ranking problem in two ways: (1) using the LTR (Learning to Rank) technique, (2) training a binary classifier. In approach (1), a ranking model is learnt by treating the training data has having valued ranks — true keyphrases are assigned rank value 1 while all other candidate phrases are assigned rank value 0. During prediction, the scores predicted by the ranking model are used to sort the candidate phrases and select the top k. In approach (2), we directly train a binary classifier; during prediction, the positive class's probability score in the classifier's output is treated as the ranking score.

### 3.2. Features

We describe below the features we compute for each candidate phrase and pass as inputs to the model.

#### 3.2.1. Statistical features

These features represent various distributional properties of a phrase in the context of the document where it appears and/or with respect to the collection of documents seen in the training set. A noteworthy and distinguishing aspect of our work is that unlike other known approaches using statistical features, we do not use TF-IDF as a feature. We decouple term frequency from document frequency for reasons that are elaborated on in Appendix; we believe that this decoupling enables our model to achieve a higher accuracy than what it would have had TF-IDF been used instead. Our formulation of document frequency too differs slightly from conventional definitions of document frequency in TF-IDF.

**Phrase count:** The number of times a candidate phrase appears in the document. This feature is motivated by the empirical observation that phrases that occur just once or twice in a document are less likely to be central to the main topics or entities of the content. Keyphrases usually occur multiple times in the document, though not necessarily having to be among say the top 5 most frequent phrases. Very high frequency phrases may turn out to be non-keyphrases (for e.g., "legal process" can be a very frequent term in legal documents but would generally not be considered a keyphrase of that document), but this depends a lot on the content and presentation of a document. It is the responsibility of our model to infer the relationship between frequency and the likelihood of a phrase being a keyphrase.

**Document frequency (max-scaled):** The scaled document frequency of a candidate phrase is computed as:

$$DFS(p_D) = \frac{|\{d \in T : p_D \in d\}|}{max_{q \in d', d' \in T}|\{d \in T : q \in d\}|}$$

where $p_D$ is a candidate phrase for an input document $D$ and $T$ is the set of documents in the training set. This feature is driven by the widely observed trend that terms that appear in a very high proportion of documents are often too generic to be a keyphrase. For e.g., "related work" and "references" are seen in most research papers and these terms are seldom keyphrases. Note that we do not follow the convention of applying logarithmic transformation after scaling; instead we let the model implicitly learn the optimal transformation.

**Suffix phrase frequency:** This feature is based on the fact that a basal noun phrase may be used with different prefix adjectives in different sentences within a document. Currently we consider only the last 2 words of a phrase as the suffix. For example, in an article having "graph colouring" as its main topic, the phrase may occur as a complete (without adjectives) noun phrase in a few sentences whereas in other sentences it may appear as a part of a larger noun phrases like "conventional graph colouring" and "approximate graph colouring". Longer N-grams (i.e. larger values of N) generally have a lower frequency than shorter ones. Knowing that the 3-gram "approximate graph colouring" has a suffix that appears quite frequently within the document despite the 3-gram itself occurring just once or twice in a document could boost the prospects of that 3-gram being a good keyphrase candidate. Note: For 1-gram and 2-gram phrases, this feature defaults to the full phrase's frequency.

**Suffix phrase document frequency:** This is the document frequency (scaled variant) counterpart of the suffix phrase frequency, analogous to how the earlier mentioned document frequency feature complements term frequency. With reference to the previous example, "approximate graph colouring", this feature conveys the number of documents (from training corpus) in which the 2-gram suffix, i.e. "graph colouring" occurs.

**Suffix phrase average per-doc frequency:** The average number of times the 2-gram suffix $p_s$ of a given candidate phrase $p$ occurs as a complete phrase in training documents where $p_s$ appears at least once, i.e.

$$\frac{\sum_{d \in T} TermFreq(p_s, d)}{|\{d \in T : p_s \in d\}|}$$

where $T$ denotes the training set documents, and $TermFreq(q, d)$ is the term frequency (number of occurrences) of phrase $q$ in document $d$. The idea behind this feature is that phrases occurring more frequently in the current document than in other documents may tend to have a higher chance of being keyphrases for the current document.

**Word combination likelihood:** Given a phrase $p = w_1\ w_2\ \ldots\ w_n$ where $w_1, w_2, \ldots, w_n$ are the component words (unigrams), this feature is computed as:

$$\sqrt[n]{\prod f_{pref}(w_1) f_{pref}(w_2) \cdots f_{pref}(w_{n-1}) f_{suff}(w_n)}$$

where $f_{pref}(w_i)$ denotes the document frequency (max-scaled) of $w_i$ as a suffix, i.e. indicative of the number of documents from the training corpus where $w_i$ appears as the prefix (first word) of some phrase, and $f_{suff}(w_i)$ analogously represents the number of documents where $w_i$ appears as the suffix (last word) of some phrase. This feature serves as a rough estimate of the likelihood of a multigram phrase occurring by random chance combination of its component words. Multigrams generally have lower frequencies than unigrams. The geometric mean-based estimate is an effort to make the model more robust to the such scale differences in frequencies and also to help it learn how the difference between random chance likelihood and the actual document frequency of a candidate phrase influences the choice of keyphrases.

### 3.2.2. Positional and other features

**First occurrence index:** Index of first occurrence of a given phrase in the list of all candidate phrases extracted from the document, scaled down and rounded down. We chose this feature because the benchmark datasets are comprised of documents that are academic papers or reports which have a summary or abstract section at the beginning, followed by an introduction-like section. In such structured document, the position of first occurrence of a phrase bears a strong correlation with the likelihood of it being a keyphrase. Many documents were found to have at least one or two keyphrases taken from the abstract. To avoid overfitting, we divide the index by a constant (set to 25 in our experiments) and round down the result to nearest integer.

**N-gram size:** Number of words in a given phrase. Used since long (more than 4 words in case of benchmark data) phrases are rarely seen in manually assigned keyphrases, and among short phrases too, there could be a preference for say 2-grams over 3-grams or vice-versa.

### 3.3. Implementation Details

We use the SpaCy library [13] for the candidate phrase extraction. We select only noun phrases consisting of no more than one adjective followed by one or more nouns; our prior analysis of multiple benchmark datasets showed that most keyphrases are noun phrases of the mentioned type. Noun phrases containing more than one adjective prefix are trimmed to retain just one adjective immediately before the noun(s) in the phrase. Unlike other algorithms for AKE, we do not perform stemming on the phrases; instead we lemmatize the base noun in the noun phrase to convert plural form to singular form. During the training phase, the extracted phrases are first aggregated across documents and written to a phrase document frequency corpus that is subsequently used for model training and well as for prediction for new documents.

Since the datasets we used to train and evaluate our model contained only academic papers, though from various subjects, during the candidate extraction step we chose to skip text present in "References" and "Acknowledgements" sections. This was done with the intention of reducing noise in the data, as we anticipated present keyphrases to be present in the main content of the document. To our surprise, during model evaluation we found a few articles where some keyphrases were concentrated in the references section. However, we decided to treat these as exceptional cases and chose not to change the exclusion logic. Except for this simple exclusion logic, we do not use the section type anywhere. We want our model to generalize well across document types — having additional filtering logic or features based on section types would work against such generalization ability.

As mentioned earlier, we try two approaches to rank candidates. We train an XGBRanker (gradient boosting based ranker, [14]) model for direct ranking method and XGBClassifier (gradient boosting based classifier, [14]) for classification-based method. In both cases, the training label is binary valued with 1 indicating that the candidate phrase is a keyphrase for that document and 0 indicating otherwise, The XGBRanker model expects the training set to specify a grouping of rows based on the "query" to which those rows pertains, the idea being that when the ranker is trained, it must compare only objects that would actually compete with one another for getting ranked. Since candidate phrases are to be ranked within the scope of the single document to which all of them belong, this is easily achieved by setting the "qid" column (see XGBRanker documentation) to the unique id of the document as per the dataset.

## 4. EVALUATION

We evaluate our model on two benchmark datasets of English documents — SemEval2010, Krapivin. Table 1 presents a statistical overview of each dataset. We use F1-score as the metric with which to compare our model with benchmark models. More specifically, we compute the

F1 scores achieved when the top 5 and top 10 keyphrases as per the model are selected (i.e., F1@5 and F1@10). The results are shown in Table 2. We did not run experiments to evaluate the benchmark models. Instead, we present the F1-scores as reported by the original papers of those models or by the survey papers we referenced [2, 3].

Table 1. Overview of the datasets (#Doc: number of documents, Avg. KP: average number of keyphrases per document, Absent KP%: percentage of absent keyphrases)

| Dataset | #Doc | Avg. words per doc | Avg. KP per doc | Absent KP% |
|---|---|---|---|---|
| Krapivin | 2304 | 8040 | 6.34 | 15.3% |
| SemEval2010 | 244 | 8332 | 16.47 | 11.3% |

Table 2. Performance of our model compared with some well-known benchmark models. Some scores are left blank for some benchmark models due to unavailability of complete evaluation results for those in referenced papers

| | | Krapivin | | SemEval2010 | |
|---|---|---|---|---|---|
| Model | Nature of the model | F1@5 | F1@10 | F1@5 | F1@10 |
| TF-IDF | Unsupervised | 0.115 | 0.140 | 0.161 | 0.167 |
| TextRank | Unsupervised; graph-based | 0.148 | 0.139 | 0.168 | 0.183 |
| ExpandRank | Unsupervised; graph-based | 0.096 | 0.136 | 0.135 | 0.163 |
| Maui | Supervised | 0.249 | 0.216 | 0.178 | 0.172 |
| WINGNUS | Supervised | - | - | 0.205 | 0.247 |
| EmbedRank | Unsupervised; Deep learning-based | 0.131 | 0.138 | 0.108 | 0.105 |
| AutoKeyGen | Unsupervised; Deep learning-based (Seq2Seq) | 0.171 | 0.155 | 0.187 | 0.240 |
| UKERank | Unsupervised; Deep learning-based (BERT) | - | - | 0.180 | 0.254 |
| CopyRNN | Supervised; Deep learning-based | 0.302 | 0.252 | 0.291 | 0.296 |
| CatSeq | Supervised; LLM-based | 0.307 | 0.274 | 0.302 | 0.306 |
| KeyBART | Supervised; LLM-based | 0.292 | - | 0.274 | - |
| XGBRanking using NDCG (our model) | Supervised | 0.211 | 0.225 | 0.185 | 0.269 |
| XGBRanking using MAP (our model) | Supervised | 0.213 | 0.228 | 0.186 | 0.268 |
| XGBClassif (our model) | Supervised | 0.245 | 0.255 | 0.171 | 0.258 |

For the direct ranking variant of our model (denoted by "XGBRanking" in Table 2), we try two subvariants that differ based on the objective function used when training the ranker — one version of the model uses Maximum Average Precision (MAP) while the other uses Normalized Discounted Cumulative Gain (NDCG). Our classifier-based ranking method is denoted by "XGBClassif" in Table 2. In this method, the probability score associated with the positive class in XGBClassifier's prediction output is used as the score with which to rank the candidates.

Overall, we don't see a significant difference in terms of accuracy between the XGBClassifier-based method and the XGBRanker-based method, though the former performs better on Krapivin dataset and the latter performs better on SemEval2010 dataset. With respect to

XGBRanker, using MAP as the objective function leads to a slightly better accuracy than NDCG.

It can be seen that all our model variants outperform the accuracy levels of all unsupervised benchmark models, including all unsupervised deep learning-based models. In fact, when considering only F1@10 scores, our model is competitive with CopyRNN, a supervised DL-based keyphrase generation model, and with CatSeq, an LLM-based model. The F1@10 score of our XGBClassifier model variant betters that of CopyRNN on Krapivin dataset while in SemEval2010 dataset, the score is less than CopyRNN's by only 0.038. Admittedly, the F1@5 scores of our model are significantly lower than those of supervised DL models on SemEval2010 dataset. This may be related to the fact that the average number of present keyphrases per document in SemEval2010 dataset is about 14 whereas is it only 5 in case of Krapivin dataset. Our model perhaps has a higher level of confusion in selecting the top 5 candidates but performs better as the selection size increases with most of the true top 5 keyphrases making their way to the top 10 selected by the model. A more plausible explanation is the higher level of subjectivity and inconsistency seen in the choice of true keywords in SemEval2010 dataset. For example, document J-10 in that dataset has two 1-gram keyphrases in the truth set, namely "rating" and "correlation", having TF-IDF scores 225.84 and 8.705 respectively. But the document also contains candidate keywords "cleanliness" and "tripadvisor" having TF-IDF scores 35.75 and 42.9 respectively, i.e. significantly higher than the score of "correlation". And these two words also occur earlier in the document than the first occurrence of "correlation". Yet, neither of these two candidate keywords are present in the true keyphrase list for J-10. In a few other documents in SemEval2010, we see a more consistent correlation between the term frequency and/or first occurrence on one hand and the likelihood of the phrase being a true keyphrase. In contrast, in most documents in Krapivin dataset, the choice of true keyphrases bears a more consistent correlation with our features, i.e. there is less noise due to subjectivity in keyphrase selection. This obviously makes its easier for the model to learn patterns undelying the choice of keyphrases in Krapivin dataset.

Another observation that might raise questions is that Maui and WINGNUS, two non-DL supervised models, achieve F1@5 scores very close to or marginally better than those of our model. However, this can be explained by the fact that those models make use of external knowledge corpora that boost their ability to perform specifically well on scholarly articles that constitute the two benchmark datasets. Maui [8] relies on a corpus of "golden" keywords for its keyphraseness feature, in effect remembering previously seen keyphrases and thereby getting a strong external hint on what phrases are more likely to be selected as keyphrases again. This will severely impede the model's ability to perform well on documents whose main topics are very different from those seen in the training set. Besides, obtaining a sufficiently large "golden" keyphrase corpus is impractical when dealing with documents coming from sources where good quality annotated data is hard or costly, for e.g., tech blogs, news articles. WINGNUS [15] makes use of DBLP, a database containing bibliographic information about millions of academic articles and conference papers in the field of computer science. It also uses features that are heavily tailor-made for academic articles, such as whether a candidate phrase appears in abstract, in references, in introduction section. Thus, WINGNUS too is unlikely to perform well on documents that are not scholarly articles, or even academic articles from outside the computer science field. In contrast, our model does not rely on any annotated data during prediction phase. Though we make use of a phrase document frequency corpus, this corpus can be created from non-annotated documents and is also not constrained to be created from academic articles. Thus, our model can be expected to generalize better than Maui and WINGNUS on new types of documents.

## 5. CONCLUSION AND FUTURE WORK

We presented a novel keyphrase extraction technique that is advantageous in various angles compared to prior work. Our method uses a lightweight ensemble model and yet achieves accuracy levels significantly better than most of the benchmark models and quite close to the levels achieved by state-of-the-art deep learning based solutions. It does not require any external knowledge base or massive annotated data to achieve these accuracy levels; it is therefore not domain-specific and is poised to generalize better. It does not need any pre-trained language models or embeddings. Another notable benefit is that our model needed to be trained only on the small to moderate sized datasets to achieve accuracy levels comparable to those of supervised DL-based models. This is in sharp contrast with supervised deep learning models that need much larger labelled training sets; CopyRNN [12], for example, was trained using more than 500000 author-annotated scientific papers.

Future work would include investigating ways to improve our model's accuracy through the introduction of some more simple features and evaluating our model on more datasets. We will also explore the utility of adding lightweight graph-based features, though it may also be noted that a few of the already used statistical features such as suffix phrase frequency implicitly capture a small number of properties that are more directly derivable as graph-based properties.

## APPENDIX

### A.1. How good is TF-IDF as an indicator of keyphraseness?

Keyphrase selection using TF-IDF (ranking by TF-IDF score and selecting the highest scoring k phrases) is almost always included as a baseline when evaluating new keyphrase extraction algorithms. This is due to the fact that despite its simplicity, TF-IDF, as well as its extensions such as BM25 [16], performs remarkably well in many information retrieval applications such as document similarity and query-to-document relevance scoring. However, in the context of keyphrase extraction we find that TF-IDF performs rather poorly and somewhat erratically too. Figures 1 and 2 show the recall levels achieved on Krapivin and SemEval datasets when using TF-IDF score alone for keyphrase selection. For the purpose of illustrating the effect clearly, we plot the recall of multigram keyphrases alone (unigrams often have much higher TF-IDF scores than mutiwords phrases, so the effect of TF-IDF will be less obvious if we plotted the recall considering all keyphrases).

It can be seen from Figure 1 that in Krapivin dataset, TF-IDF barely achieves a recall of more than 0.2 even when up to 10 candidate keyphrases are selected for documents having 3 to 5 true keyphrases. The recall drops significantly for documents having more than 6 true keyphrases. In the SemEval dataset too (Figure 2), the recall barely crosses 0.2 when up to 10 keyphrases are selected. The recall is also more erratic in this dataset as it can be observed that the recall levels differ a lot between documents having different numbers of "true" (expected) keyphrases. This erratic behaviour is more prominent when the number of keyphrases selected is less than 8 whereas the recall plateaus out at around 0.2. Interestingly, a sharp increase in recall is seen for documents having 5 true keyphrases when the number of candidates selected is increased from 6 to 8, but this is also a reflection of the more subjective and more inconsistent selection criteria used for true keyphrases in SemEval dataset.

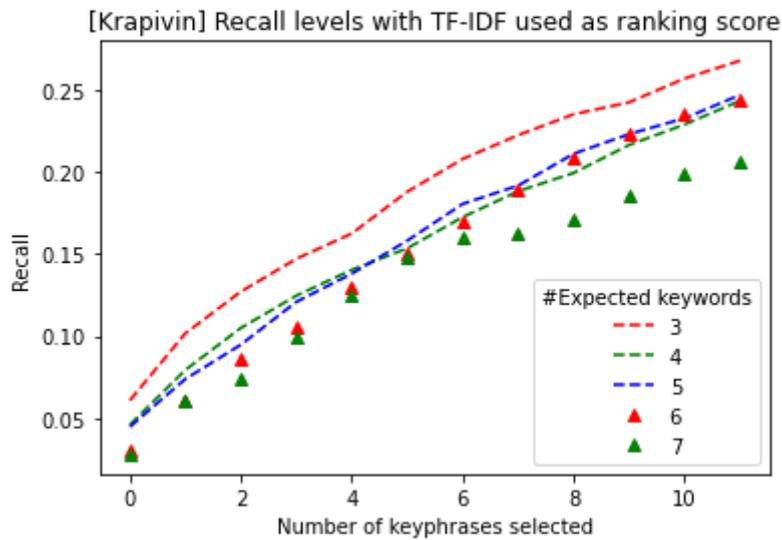

Figure 1. Recall of multigram keyphrases for Krapivin dataset when TF-IDF is used as the scoring function. Multiple plots are used to illustrate the recall for different documents grouped by the number of expected ("true") keyphrases

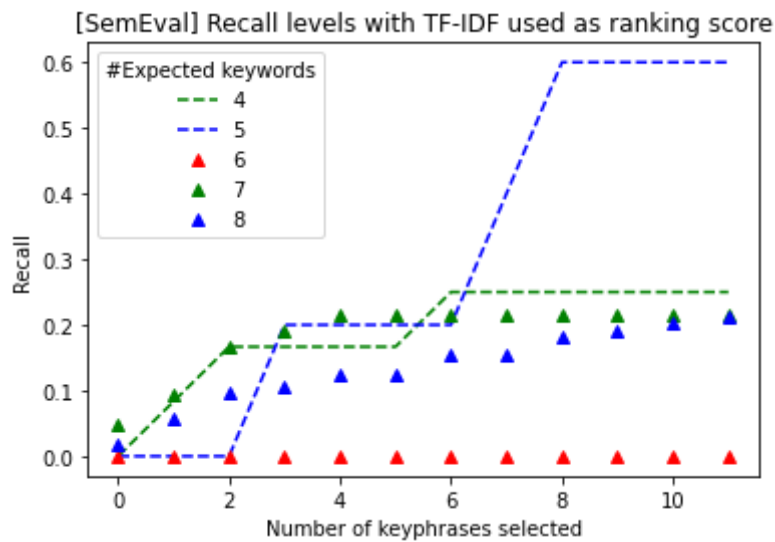

Figure 2. Recall of multigram keyphrases for SemEval2010 dataset when TF-IDF is used as the scoring function. Multiple plots are used to illustrate the recall for different documents grouped by the number of expected (true) keyphrases

In summary, while TF-IDF has a positive correlation with keyphraseness (i.e. likelihood of a candidate phrase being a true keyphrase), the correlation is rather weak. Besides, this statistic cannot adapt itself when the underlying selection criteria are based on properties other than term and document frequencies of phrases. Thus, there is no strong reason to support the use of TF-IDF, which in itself is a heuristically derived statistic, as a feature in a classifier. Instead, we could let the classifier learn how to balance between term frequency and document frequency in the light of other properties of candidate phrases.